\documentclass[jor, amsmath, amssymb, preprint, floatfix, unsortedaddress
]{revtex4-2}

\usepackage{graphicx}
\usepackage{dcolumn}
\usepackage{bm}

\usepackage[utf8]{inputenc}
\usepackage[T1]{fontenc}
\usepackage{mathptmx}
\usepackage{etoolbox}

\usepackage{multirow}
\usepackage{enumerate}

\usepackage[dvipsnames]{xcolor}

\newcommand{\Gp}{G^{\prime}(\omega)}
\newcommand{\Gpp}{G^{\prime\prime}(\omega)}

\newcommand{\Gst}{G^{*}(\omega)}

\newcommand{\eTst}{\eta^{*}_{3}}

\begin{document}


\title[Kramers-Kronig for Nonlinear Rheology]{Kramers-Kronig Relations for Nonlinear Rheology: 1. General Expression and Implications\footnote{This paper is dedicated to the memory of Prof. James W. Swan whose contributions inspired this work.}}

\author{Sachin Shanbhag}%
 \email{sshanbhag@fsu.edu}
\affiliation{Department of Scientific Computing, Florida State University, Tallahassee, FL 32306. USA}

\author{Yogesh M. Joshi}%
\email{joshi@iitk.ac.in}
\affiliation{Department of Chemical Engineering, Indian Institute of Technology, Kanpur, INDIA}

\date{\today}

\begin{abstract}
The principle of causality leads to linear Kramers-Kronig relations (KKR) that relate the real and imaginary parts of the complex modulus $G^{*}$ through integral transforms. Using the multiple integral generalization of the Boltzmann superposition principle for nonlinear rheology, and the principle of causality, we derived nonlinear KKR, which relate the real and imaginary parts of the $n^\text{th}$ order complex modulus $G_{n}^{*}$. For $n$=3, we obtained nonlinear KKR for medium amplitude parallel superposition (MAPS) rheology. A special case of MAPS is medium amplitude oscillatory shear (MAOS); we obtained MAOS KKR for the third-harmonic MAOS modulus $G_{33}^{*}$; however, no such KKR exists for the first harmonic MAOS modulus $G_{31}^{*}$. We verified MAPS and MAOS KKR for the single mode Giesekus model. We also probed the sensitivity of MAOS KKR when the domain of integration is truncated to a finite frequency window. We found that that (i) inferring $G_{33}^{\prime\prime}$ from $G_{33}^{\prime}$ is more reliable than vice-versa, (ii) predictions over a particular frequency range require approximately an excess of one decade of data beyond the frequency range of prediction, and (iii) $G_{33}^{\prime}$ is particularly susceptible to errors at large frequencies.
\end{abstract}


\maketitle

\section{Introduction}

In the linear viscoelastic (LVE) regime, soft materials are subjected to infinitesimal deformations so that the mechanical response may be probed without perturbing the equilibrium microstructure. Linear viscoelasticity is commonly examined using three types of experiments: (i) step strain, (ii) creep or step stress, and (iii) oscillatory strain.  These experiments lead, respectively, to the linear response functions: (i) stress relaxation modulus $G(t)$, (ii) creep compliance $J(t)$, and (iii) complex modulus $\Gst$. Here, $t$ and $\omega$ represent time and frequency, respectively. These LVE response functions provide insights into material microstructure, and are part of standard rheological characterization \cite{Pipkin1972book, TschoeglPhenomenological, Ferry1980, Cho2016}. These response functions are interrelated. If complete knowledge of any one property is available, the remaining two can be inferred. The relaxation modulus $G(t)$ and creep compliance $J(t)$ are related to each other by the convolution relation. The complex modulus $\Gst$, on the other hand, is related to the Fourier transform of $G(t)$. The complex modulus has real and imaginary parts, $\Gst = \Gp + i \Gpp$, called the storage (or elastic) and loss (or viscous) modulus, respectively. These  moduli are related to each other via Kramers-Kronig relations (KKR) \cite{Kramers1929, L.Kronig1926}. They can be experimentally measured by performing oscillatory shear (OS) experiments. 

\subsection{Linear Kramers-Kronig Relations}

Suppose a material is subjected to an arbitrary time-dependent shear strain $\gamma(t)$. In the LVE regime, the induced shear stress $\sigma(t)$ is related to the deformation history by the Boltzmann superposition principle \cite{TschoeglPhenomenological, Ferry1980, Cho2016},
\begin{equation}
\sigma (t)=\int_{-\infty }^{t} G( t-t' )\dfrac{d\gamma(t')}{dt'}\,dt',
\label{eqn:boltzmann}
\end{equation}
where $t$ is the current time, and ${t}'$ is the time associated with application of deformation field. The principle of causality stipulates that the relaxation modulus $G(t - t' ) = 0$ for $t - t'<0$.  Using this principle, the complex modulus can be related to the relaxation modulus via a modified Fourier transform,
\begin{equation}
G^{*}(\omega) =  i\omega \int_{-\infty}^{\infty } G(t)\, e^{-i\omega t}\, dt = i\omega \int_{0}^{\infty } G(t)\, e^{-i\omega t} dt.
\label{eqn:ft}
\end{equation}
Note that the complex viscosity $\eta^{*}(\omega) = \eta^{\prime}(\omega) - i \eta^{\prime \prime}(\omega) = \Gst/(i \omega)$ and the relaxation modulus $G(t)$ form a standard Fourier transform pair. 

Using the principle of causality again to write $G(t)$ as a sum of an even and odd function, it can be shown that \cite{Ferry1980, Cho2016},
\begin{equation}
G(t) = \dfrac{2}{\pi }\int_{0}^{\infty } \dfrac{\Gp}{\omega} \sin \omega t \, d\omega = \dfrac{2}{\pi }\int_{0}^{\infty } \dfrac{\Gpp}{\omega } \cos \omega t \, d\omega.
\label{eqn:ift}
\end{equation}
This constrains the functional forms of $\Gp$ and $\Gpp$ to be even and odd functions of $\omega$, respectively. By manipulating equations \ref{eqn:ft} and \ref{eqn:ift}, we obtain the linear KKR \footnote{For convenience and brevity, we assume that $G^{\prime}(0) = 0$ (viscoelastic liquids), in this work. Generalizations of KKR to include nonzero equilibrium modulus $G^{\prime}(0) \neq 0$ are straightforward.}:
\begin{align}
	G^{\prime}(\omega) & = -\dfrac{2 \omega^2}{\pi} \int_{0}^{\infty} \dfrac{G^{\prime\prime}(u)/u}{u^2 -  \omega^2} du \notag\\
	G^{\prime\prime}(\omega) & = \dfrac{2 \omega}{\pi} \int_{0}^{\infty}  \dfrac{G^{\prime}(u)}{u^2 - \omega^2} du.
\label{eqn:kk_saos}
\end{align}
Since the integrals have a singularity at $u = \omega$, the Cauchy principal value of the integrals is implied. Originally, KKR were proposed for specific atomic systems using physical arguments \cite{L.Kronig1926, Kramers1927, Bohren2010}, but subsequently generalized using complex analysis, and assumptions of linearity, causality, and analyticity \cite{Toll1956,King2006}. Indeed, KKR can be derived in a succinct form from the residue theorem in complex analysis with deceptive simplicity as \cite{Hu1989},
\begin{equation}
\eta^{*}(\omega) = \dfrac{i}{\pi} \int_{-\infty}^{\infty} \dfrac{\eta^{*}(u)}{u-\omega} du.
\label{eqn:kk_saos_complex}
\end{equation}
This equation is equivalent to the two relations in equation \ref{eqn:kk_saos}. Table \ref{tab:linearKKR} summarizes different forms in which linear KKR can be expressed \footnote{Note that $\eta^{*}(\omega)$ is analogous to susceptibility $\chi(\omega)$ in optics. However, unlike $\eta^{*} = \eta^{\prime} - i \eta^{\prime\prime}$,  susceptibility is defined as $\chi^{*} = \chi^{\prime} + i \chi^{\prime\prime}$. This difference in definition results in the comparable KKR, $\chi^{*}(\omega) = 1/(i \pi) \int_{-\infty}^{\infty} \dfrac{\chi^{*}(u)}{u-\omega} du.$}.

\begin{table}
\begin{tabular}{ccc}
        & \textbf{Complex Form} & \textbf{Pair Form}\\
\hline        
\multirow{2}{*}{modulus} & \multirow{2}{*}{$\dfrac{G^{*}(\omega)}{\omega} = \dfrac{i}{\pi} \int\limits_{-\infty}^{\infty} \dfrac{G^{*}(u)/u}{u-\omega} du.$} & $G^{\prime}(\omega) = -\dfrac{2 \omega^2}{\pi} \int\limits_{0}^{\infty} \dfrac{G^{\prime\prime}(u)/u}{u^2 -  \omega^2} du$\\[2ex]
& & $G^{\prime\prime}(\omega) = \dfrac{2 \omega}{\pi} \int\limits_{0}^{\infty}  \dfrac{G^{\prime}(u)}{u^2 - \omega^2} du$\\[2ex]
\multirow{2}{*}{viscosity} & \multirow{2}{*}{$\eta^{*}(\omega) = \dfrac{i}{\pi} \int\limits_{-\infty}^{\infty} \dfrac{\eta^{*}(u)}{u-\omega} du.$} & $\eta^{\prime}(\omega)  = \dfrac{2}{\pi} \int\limits_{0}^{\infty}  \dfrac{u \eta^{\prime\prime}(u)}{u^2 - \omega^2}\, du$\\[2ex]
& & $\eta^{\prime\prime}(\omega) = \dfrac{2\omega}{\pi} \int\limits_{0}^{\infty}   \dfrac{\eta^{\prime}(u)}{\omega^2 - u^2}\, du$\\
\hline
\end{tabular}
\caption{Different forms of linear Kramers-Kronig relations. \label{tab:linearKKR}}
\end{table}

It is useful to emphasize that the relationship between the real and imaginary parts of $\Gst$ implied by KKR is underpinned by the principle of causality. It is merely a mathematical reflection of the physical constraint $G(t < 0) = 0$, mediated through the Fourier transform of a real causal function. Nevertheless, it  can be practically useful. For example, in \textit{small amplitude} oscillatory shear (SAOS) experiments  a sinusoidal strain $\gamma(t) = \gamma_0 \sin \omega t$ with amplitude $\gamma_0$ and angular frequency $\omega$ is applied \cite{Ferry1980, Dealy2009}. In the LVE limit, the stress response is given by,
\begin{equation}
\sigma(t)=\sigma_\text{SAOS}(t) = \gamma_0 \left( \Gp \sin \omega t + \Gpp \cos \omega t\right).
\label{eqn:stressSAOS}
\end{equation}
Modern rheometers can measure this response, and infer $\Gp$ and $\Gpp$ over a range of frequencies in a typical frequency sweep experiment. For thermo-rheologically simple materials, this frequency range can be widened to several decades using the time-temperature superposition principle \cite{Ferry1980}. Manual shifting of individual datasets to produce master-curves of $\Gst$ can result in violation of KKR. Thus, KKR can be used for data-validation of experimentally measured $\Gst$ \cite{Winter1997, Rouleau2013}.

\subsection{MAOS and MAPS Rheology}

In the LVE limit of $\gamma_0 \rightarrow 0$, stresses induced in a material are small and harmonic. As the magnitude of the deformation is gradually increased, nonlinear viscoelastic features start to manifest, and the stress response becomes non-harmonic. In conventional large amplitude oscillatory shear (LAOS) experiments, an oscillatory strain field $\gamma(t) = \gamma_0 \sin \omega t$ similar to SAOS experiments, is applied \cite{Tee1975, Giacomin1998, Hyun2011}. For large values of amplitude $\gamma_0$, higher nonlinear modes are activated, and the stress response is often represented by a power series \cite{Pearson1982},
\begin{equation}
\sigma (t) = \sum\limits_{n \in \text{odd}}^{\infty} \sigma_{n}(t) = \sum\limits_{n \in \text{odd}}^{\infty} \sum \limits_{m \in \text{odd}}^{n} \gamma_{0}^{n} \left[ G_{nm}^{\prime}(\omega) \sin (m \omega t) + G_{nm}^{\prime \prime}(\omega) \cos (m \omega t) \right],
\label{eqn:powser}
\end{equation}
where the nonlinear complex moduli $G_{nm}^{*}(\omega) = G_{nm}^{\prime}(\omega) + iG_{nm}^{\prime\prime}(\omega)$ are functions of only frequency. The summations include only odd values of $m$ and $n$ due to the odd symmetry of shear stress with shear strain. In the medium amplitude oscillatory shear (MAOS) regime, only the weakest nonlinear modes are activated, and the stress response $\sigma(t) =  \sigma_1(t) +  \sigma_3(t) + \mathcal{O}(\gamma_0^5)$ can be truncated after the leading nonlinear term $n=3$ in equation \ref{eqn:powser} as,
\begin{align}
\sigma(t) = & \gamma_0 \left[G_{11}^{\prime} \sin \omega t +  G_{11}^{\prime\prime} \cos \omega t \right] + \notag \\
& \gamma_0^3 \left[G_{31}^{\prime} \sin \omega t +  G_{31}^{\prime\prime} \cos \omega t +G_{33}^{\prime} \sin 3\omega t +  G_{33}^{\prime\prime} \cos 3\omega t \right],
\label{eqn:sigmaMAOS}
\end{align}
where $\sigma_1 = \sigma_\text{SAOS}$ and the LVE complex modulus is given by $G^{*} = G_{11}^{\prime} + i G_{11}^{\prime\prime}$. The stress term $\sigma_3(t)$ associated with cubic power of strain is the MAOS contribution; the MAOS moduli associated with the first and third harmonic are $G^{*}_{31} = G^{\prime}_{31} + i G^{\prime\prime}_{31}$, and $G^{*}_{33} = G^{\prime}_{33} + i G^{\prime\prime}_{33}$, respectively. MAOS measurements have been used to discriminate between linear and branched polymers \cite{Hyun2009, Wagner2011, Song2016}, evaluate nanoparticle dispersion quality \cite{Lee2016, Lim2013}, droplet size dispersion in polymer blends \cite{Ock2016, Salehiyan2014}, quantify filler-matrix interactions in filled rubbers \cite{Xiong2018, Wang1998} etc. 

Experimentally, extraction of the MAOS moduli is indirect, and involves careful extrapolation. The effort and care required is significantly greater than that required in the measurement of LVE moduli $\Gp$ and $\Gpp$, in part, due to the narrow window of suitable strain amplitudes \cite{Ewoldt2013}. If $\gamma_0$ is too small, MAOS signals are too weak and difficult to measure. If $\gamma_0$ is too large, the stress response is contaminated by the contribution of  modes higher than the third harmonic. Furthermore, the optimal range of $\gamma_0$ is frequency-dependent; at low frequencies, higher strain amplitudes are necessary to ferret out MAOS signatures. In practice, the stress response is measured at multiple strain-amplitudes in the target zone, and the ``true'' MAOS moduli are extracted by extrapolation. Due to the complicated process involved, validating experimental data before interpretation is paramount. The companion paper provides a method for efficiently accomplishing this task \cite{kkr2}.

Medium amplitude parallel superposition (MAPS) can be seen as a generalization of the MAOS protocol \cite{Lennon2020}. Instead of the single-tone sinusoidal strain in MAOS, the strain waveform in MAPS consists of a superposition of three sine waves with frequencies $\omega_1$, $\omega_2$, and $\omega_3$,
\begin{equation}
\gamma_\text{MAPS}(t) = \gamma_0 \left(\sin (\omega_1 t) +  \sin (\omega_2 t) + \sin (\omega_3 t)\right).
\label{eqn:maps}
\end{equation}
This perturbation elicits a much richer asymptotic nonlinear response than MAOS. Indeed, as introduced formally in section \ref{sec:generalderivation} it leads to a strain-independent third-order complex modulus $G_3^{*}(\omega_1, \omega_2, \omega_3)$ which offers a \textit{complete} characterization of the material's asymptotic nonlinear behavior. By complete, we mean that using $G_3^{*}(\omega_1, \omega_2, \omega_3)$ the nonlinear response to any arbitrary medium amplitude deformation history can be  predicted via a generalization of Boltzmann superposition principle. 

Therefore, MAOS can be thought of as a special, low-dimensional projection of  MAPS: it can be shown that the MAOS moduli, $G_{31}^{*}$ and $G_{33}^{*}$, are special cases of the third order or MAPS modulus $G_{3}^{*}(\omega_1, \omega_2, \omega_3)$ \cite{Lennon2020},
\begin{align}
G_{31}^{*}(\omega) & = \dfrac{3}{4} G_3^{*}(\omega, -\omega, \omega) \notag\\
G_{33}^{*}(\omega) & = -\dfrac{1}{4} G_3^{*}(\omega, \omega, \omega).
\label{eqn:MAOS_G3_2}
\end{align}

Due to experimental challenges, characterization of materials using MAPS has barely started \cite{Lennon2020a, Lennon2021, Lennon2021a}. Regardless, for the purposes of this work, MAPS provides a convenient general theoretical lens for interpreting MAOS measurements and KKR.

\subsection{Organization}

This paper is organized as follows: we begin with a generalization of the Boltzmann superposition principle to nonlinear rheology using a multiple integral expansion in section \ref{sec:generalderivation}. We highlight similarities and connections between higher-order terms and their LVE counterparts. We then mathematically derive a  general nonlinear KKR (equation \ref{eqn:kk_nonlinear}). 

In section \ref{sec:kkr3rdorder}, we narrow our focus by specializing these general KKR to MAPS and MAOS rheology. It turns out that for the MAOS modulus $G_{33}^{*}$ we can formulate a KKR (equation \ref{eqn:kkr_G33}); unfortunately no such KKR exists for $G_{31}^{*}$. Finally, in section \ref{sec:applications}, we test KKR on the single mode Giesekus model for which the third-order complex modulus $G_3^{*}(\omega_1, \omega_2, \omega_3)$ is analytically known. In particular, we verify that the MAPS  moduli are consistent with the appropriate KKR. Finally, we explore the sensitivity of the MAOS KKR when the domain of integration is limited to a finite frequency window.

\section{Derivation of Kramers-Kronig Relations for Nonlinear Rheology}

\label{sec:generalderivation}

The Boltzmann superposition principle can be generalized to nonlinear rheology using a multiple integral expansion \cite{Volterra1959, Bierwirth2019, Findley1976, Davis1978, Lennon2020}. The general framework for stress induced due to imposed strain can be represented using an infinite Volterra series,
\begin{equation}
\sigma ( t ) = \sum_{n \in \text{odd}}^{\infty} \, \sigma_{n}(t),\label{eqn:VolterraSum}
\end{equation}
where the summation includes only odd values of $n$ due to the odd symmetry between shear stress and strain, i.e, $\sigma(-\gamma) = -\sigma(\gamma)$. The contribution of the $n^\text{th}$ mode is given by,
\begin{equation}
\sigma_{n}(t) = \int_{-\infty }^{t} \cdots \int_{-\infty }^{t} G_{n}( t-t_{1}, t-t_{2}, \cdots , t-t_{n} ) \prod_{m=1}^{n} \dot{\gamma } ( t_{m} )\, dt_{m},
\label{eqn:Volterra1}
\end{equation}
where $\dot{\gamma}(t)$ is the shear rate, and $G_{n}( t-t_{1}, t-t_{2}, \cdots , t-t_{n})$ is the $n^\text{th}$ order relaxation modulus that generalizes the linear relaxation modulus.  The principle of causality stipulates that $G_{n}( t-t_{1}, t-t_{2}, \cdots , t-t_{n}) = 0$ if $t - t_i < 0$ for any $i = 1, 2, \cdots, n$. The first term in this series,
\begin{equation}
\sigma_{1} = \int_{-\infty }^{t} G_{1} ( t-t_{1}) \, \dot{\gamma}(t_{1}) dt_{1}
\end{equation}
is identical to the Boltzmann superposition principle given by equation \ref{eqn:boltzmann} with $G_{1}(t) \equiv G(t)$. Subsequent terms ($n \geq 3$)
in equation \ref{eqn:Volterra1} take into account stress induced due to the interaction of strains applied at different times $t_i$ and $t_j$. In LVE, such cross-effects are negligible.

Nonlinear effects in oscillatory shear flow can evaluated by taking a Fourier transform (denoted by ``hat'') of equation \ref{eqn:VolterraSum},
\begin{equation}
\hat{\sigma} (\omega) = \sum_{n \in \text{odd}}^{\infty} \, \hat{\sigma}_{n} ( \omega ) = \sum_{n \in \text{odd}}^{\infty} \int_{-\infty }^{\infty} \sigma_{n}(t)\,  e^{-i \omega t } dt
\end{equation}
where the contribution of the $n^\text{th}$ mode is,
\begin{equation}
\hat{\sigma}_{n}(\omega) = \dfrac{1}{(2\pi)^{n-1}} \int_{-\infty}^{\infty} \overset{n}{\cdots} \,\int_{-\infty}^{\infty} G_{n}^{*}(\omega_{1}, \ldots , \omega_{n})\, \delta\left(\omega -\sum_{m=1}^{n} \omega_{m} \right) \left(\prod_{m=1}^{n} \hat{\dot{\gamma}}(\omega_{m})\, d\omega_{m}\right).
\label{eqn:VolterraFou}
\end{equation}
The nonlinear complex relaxation modulus $G_{n}^{*}\left( {\omega_{1}},{\omega_{2}},\ldots ,{\omega_{n}} \right)$ is the modified Fourier transform of the nonlinear relaxation modulus $G_{n} (t_{1}, \ldots, t_{n})$,
\begin{equation}
G_{n}^{*}(\omega_{1}, \ldots , \omega_{n}) = \left(\prod_{m=1}^{n} i \omega_m \right)\int_{0}^{\infty} \overset{n}{\cdots} \,\int_{0}^{\infty} G_n(t_1, \cdots, t_n)  \left( \prod_{m=1}^{n} e^{-i \omega_m t_m} dt_m \right).
\label{eqn:FourierNonlinear} 
\end{equation}
Note that the first term corresponding to $n = 1$ is the usual linear complex modulus $G_1^*(\omega) = \Gst$ encountered in the LVE regime as equation \ref{eqn:ft}. The next odd term corresponding to $n = 3$, or $G_3^{*}(\omega_1, \omega_2, \omega_3)$, represents the leading nonlinear term that can be experimentally characterized using MAPS rheology. Due to the Volterra representation, $G_n^{*}$ obeys permutation symmetry and is invariant with respect to the permutation of its arguments. Thus, for example, $G_3^{*}(\omega_1, \omega_2, \omega_3) = G_3^{*}(\omega_2, \omega_1, \omega_3) = G_3^{*}(\omega_3, \omega_2, \omega_1)$, etc.

Analogous to the linear complex viscosity $\eta^{*}(\omega) = G^{*}(\omega)/(i \omega)$, it is convenient to introduce the $n^\text{th}$ order complex viscosity $\eta_{n}^{*}(\omega_1, \cdots, \omega_n) = \eta_n^{\prime}(\omega_1, \cdots, \omega_n) - i\eta_n^{\prime\prime}(\omega_1, \cdots, \omega_n)$, which is related to the $n^\text{th}$ order complex modulus $G_{n}^{*}$ via,
\begin{equation}
G_{n}^{*}(\omega_1, \cdots, \omega_n) = \left(\prod_{m=1}^{n} i \omega_j \right)  \eta_{n}^{*}(\omega_1, \cdots, \omega_n).
\label{eqn:GstEtaSt}
\end{equation}
Using this definition, we can write equation \ref{eqn:FourierNonlinear} as,
\begin{equation}
\eta^{*}_{n}(\omega_1, \cdots, \omega_n) = \int_{0}^{\infty} \overset{n}{\cdots} \,\int_{0}^{\infty} G_n(t_1, \cdots, t_n)  \left( \prod_{m=1}^{n} e^{-i \omega_m t_m} dt_m \right).
\label{eqn:etanst}
\end{equation}

Now that we have defined all the relevant terms, we can begin deriving a general form of KKR following the approach of Hutchings et al. \cite{Hutchings1992}. Let $u$ and $\omega$ denote arbitrary frequencies. Consider the following integral with $p_i \geq 0$ for all $i = 1, \cdots, n$,
\begin{equation}
I  =  \int_{-\infty}^{\infty} \dfrac{\eta^{*}_{n}(\omega_1 + p_1 u, \cdots, \omega_n + p_n u)}{u - \omega} du
\label{eqn:IntegralConsider}
\end{equation}
Substituting equation \ref{eqn:etanst} for $\eta_{n}^{*}$, 
\begin{equation}
I =  \int_{-\infty}^{\infty} \int_{0}^{\infty} \overset{n}{\cdots} \,\int_{0}^{\infty} G_n(t_1, \cdots, t_n)\, e^{-i \sum \omega_m t_m} \cdot \dfrac{e^{-i u \sum p_m t_m}}{u - \omega}\,  \left( \prod_{m=1}^{n} dt_m \right) du,
\end{equation}
where the summations that occur as arguments to exponential functions, $\sum \omega_m t_m$ and $\sum p_m t_m$, run from $m = 1$ to $n$. We can switch the order of integration to isolate terms that involve $u$ and obtain,
\begin{equation}
I  = \int_{0}^{\infty} \overset{n}{\cdots} \,\int_{0}^{\infty} G_n(t_1, \cdots, t_n)\, e^{-i \sum \omega_m t_m} \left( \prod_{m=1}^{n} dt_m \right) \int_{-\infty}^{\infty} \dfrac{e^{-i u \sum p_m t_m}}{u - \omega} du
\label{eqn:derive_int1}
\end{equation}
We can analytically compute the integral over $u$, by using the following result for constant $a \ne 0$,
\begin{equation}
\int_{-\infty}^{\infty} \dfrac{e^{-i u a}}{u - \omega} du = (-i \pi) e^{-i \omega a}.
\label{eqn:integral_result}
\end{equation}
Using equation \ref{eqn:integral_result} in equation \ref{eqn:derive_int1}, and invoking equation \ref{eqn:etanst} in the last step, we can show that,
\begin{align}
I & =  (-i \pi) \int_{0}^{\infty} \overset{n}{\cdots} \,\int_{0}^{\infty} G_n(t_1, \cdots, t_n) e^{-i \sum \omega_m t_m} \cdot e^{-i \omega \sum p_m t_m} \left( \prod_{m=1}^{n} dt_m \right) \notag\\
& = (-i \pi) \int_{0}^{\infty} \overset{n}{\cdots} \,\int_{0}^{\infty} G_n(t_1, \cdots, t_n) e^{-i \sum (\omega_m + p_m \omega) t_m} \left( \prod_{m=1}^{n} dt_m \right) \notag\\
& = (-i \pi) \,  \eta^{*}_{n}(\omega_1 + p_1 \omega, \cdots, \omega_n + p_n \omega).
\label{eqn:IntegralEquiv}
\end{align}
We can equate the RHS of equations \ref{eqn:IntegralConsider} and \ref{eqn:IntegralEquiv} to obtain a general form of nonlinear KKR,
\begin{equation}
\eta^{*}_{n}(\omega_1 + p_1 \omega, \cdots, \omega_n + p_n \omega) = \dfrac{i}{\pi}  \int_{-\infty}^{\infty} \dfrac{\eta_{n}^{*}(\omega_1 + p_1 u, \cdots, \omega_n + p_n u)}{u - \omega} du.
\label{eqn:kk_nonlinear}
\end{equation} 
Note that this relation holds when $p_i \geq 0$ for all $i = 1, \cdots, n$, with at least one $p_i > 0$, due to equation \ref{eqn:integral_result}. This is the most general form of KKR for nonlinear rheology in this work. Several other useful forms are special cases of this relation. A particular special case is obtained by substituting $p_j = 1$ and $p_i = 0$ for all $i \ne j$ where $1 \leq i, j \leq n$, $\omega_j = 0$, $u = \omega_j^{\prime}$, and $\omega = \omega_j$,
\begin{equation}
\eta^{*}_{n}(\omega_1, \cdots , \omega_j, \cdots, \omega_n) = \dfrac{i}{\pi}  \int_{-\infty}^{\infty} \dfrac{\eta^{*}_{n}(\omega_1, \cdots , \omega_j^{\prime}, \cdots, \omega_n)}{\omega_j^{\prime} - \omega_j} d \omega_j^{\prime}.
\label{eqn:kk_nonlinear_eta}
\end{equation}
Note that the RHS involves integrating over the $j^\text{th}$ input frequency.  For $n=1$, the correspondence with the linear KKR in equation \ref{eqn:kk_saos_complex} is obvious. We can obtain an equivalent form in terms of higher order complex modulus by using equation \ref{eqn:GstEtaSt},
\begin{equation}
\dfrac{G^{*}_{n}(\omega_1, \cdots , \omega_j, \cdots, \omega_n)}{\omega_j} = \dfrac{i}{\pi}  \int_{-\infty}^{\infty} \dfrac{G^{*}_{n}(\omega_1, \cdots , \omega_j^{\prime}, \cdots, \omega_n)/\omega_j^{\prime}}{\omega_j^{\prime} - \omega_j} d \omega_j^{\prime}.
\label{eqn:kk_nonlinear_Gst}
\end{equation}

\section{Kramers-Kronig Relations for MAPS and MAOS}
\label{sec:kkr3rdorder}

\begin{figure}
    \centering
    \includegraphics[scale=0.5]{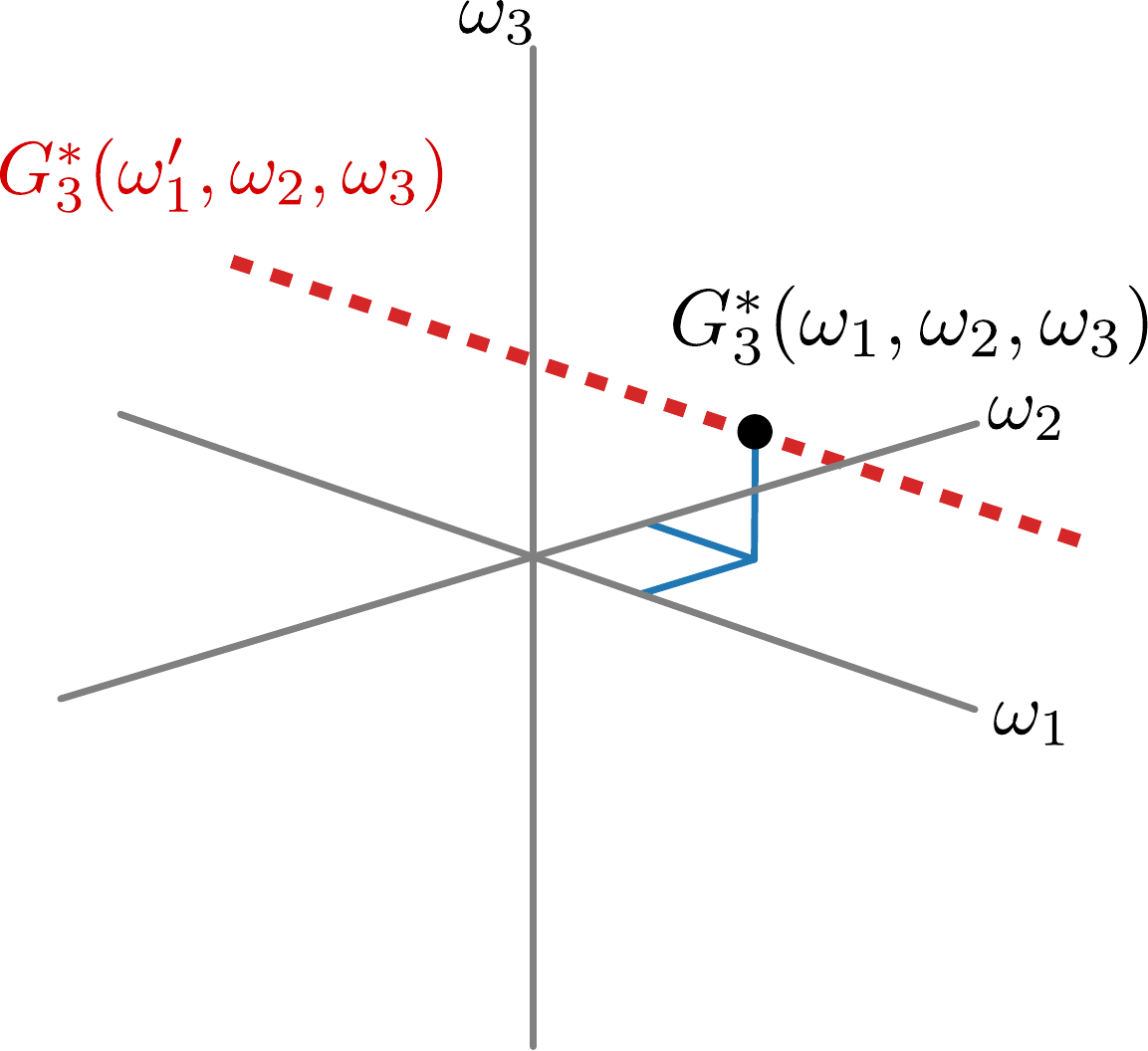}
    \caption{KKR for MAPS relates the third-order modulus at a point (black circle) to an integral over the first input frequency ($\omega_1^{\prime}$) (see equation \ref{eqn:kk_maps}) illustrated by the dashed red line.}
    \label{fig:schematic_maps}
\end{figure}

We can specialize the general forms of KKR derived for nonlinear rheology (equations \ref{eqn:kk_nonlinear} - \ref{eqn:kk_nonlinear_Gst}) for MAPS and MAOS experiments. For MAPS, two useful forms follow directly from equations \ref{eqn:kk_nonlinear_eta} and \ref{eqn:kk_nonlinear_Gst} with $n = 3$, and (say) $j = 2$:
\begin{align}
\eTst(\omega_1, \omega_2, \omega_3) & = \dfrac{i}{\pi}  \int_{-\infty}^{\infty} \dfrac{\eTst(\omega_1, \omega_2^{\prime}, \omega_3)}{\omega_2^{\prime} - \omega_2} d\omega_2^{\prime}. \notag\\
\dfrac{G^{*}_{3}(\omega_1, \omega_2, \omega_3)}{\omega_2} & = \dfrac{i}{\pi}  \int_{-\infty}^{\infty} \dfrac{G^{*}_{3}(\omega_1, \omega_2^{\prime}, \omega_3)/\omega_j^{\prime}}{\omega_2^{\prime} - \omega_2} d \omega_2^{\prime}.
\label{eqn:kk_maps}
\end{align}
For concreteness, the RHS of the equations above involve integrating over the second input frequency. Due to permutation symmetry, equivalent relations can also be furnished for first and third input frequencies. Figure \ref{fig:schematic_maps} illustrates this relation for the case where the integral is expressed over the first input frequency. This family of KKR is useful for validating MAPS experiments where  $G_{3}^{*}(\omega_1, \omega_2, \omega_3)$ or $\eta_{3}^{*}(\omega_1, \omega_2, \omega_3)$ is available.

\subsection{KKR for MAOS}

We can manipulate the general KKR relation (equation \ref{eqn:kk_nonlinear}) to develop KKR that are useful for relating the real and imaginary parts of the MAOS moduli $G_{33}^{*}$, where the perturbation is single-tone. With $n = 3$, we set $\omega_1 = \omega_2 = \omega_3 = 0$, and $p_1 = p_2 = p_3 = 1$ to obtain,
\begin{equation}
\eTst(\omega, \omega, \omega) = \dfrac{i}{\pi}  \int_{-\infty}^{\infty} \dfrac{\eTst(u, u, u)}{u - \omega} du.
\label{eqn:kkr_eta33}
\end{equation}
Using equation \ref{eqn:MAOS_G3_2} and \ref{eqn:GstEtaSt}, we can rewrite the corresponding KKR in terms of the modulus,
\begin{equation}
G_{33}^{*}(\omega) = \dfrac{i}{\pi} \omega^3 \int_{-\infty}^{\infty} \dfrac{1}{u^3} \dfrac{G_{33}^{*}(u)}{u - \omega} du.
\label{eqn:kkr_G33}
\end{equation}
We can equate the real and imaginary parts separately to obtain a pair of KKR relations. Using $G_{33}^{\prime}(-\omega) = G_{33}^{\prime}(\omega)$ and $G_{33}^{\prime\prime}(-\omega) = -G_{33}^{\prime\prime}(\omega)$, we can express these KKR  for MAOS moduli $G_{33}^{*}$ on a non-negative frequency domain similar to the linear KKR as,
\begin{align}
G_{33}^{\prime }(\omega) & = -\dfrac{2\omega^{4}}{\pi} \int_{0}^{\infty} \dfrac{1}{u^{3}} \dfrac{G_{33}^{\prime\prime}(u)}{\left(u^{2}- \omega^{2} \right)}du \notag\\
G_{33}^{\prime \prime }(\omega) & = \dfrac{2\omega^{3}}{\pi} \int_{0}^{\infty }\dfrac{1}{{{u}^{2}}}\dfrac{G_{33}^{\prime }(u)}{\left( u^{2}- \omega^{2} \right)}du.	\label{eqn:kk_maos}
\end{align}
Just like linear KKR, these MAOS KKR can be used for numerically evaluating one signal from the other, or for data validation. Similar relations for the third-harmonic are widely used in nonlinear optics \cite{Hutchings1992, Peiponen2004, Boyd2008}.

\begin{figure}
    \centering
    \includegraphics[scale=0.5]{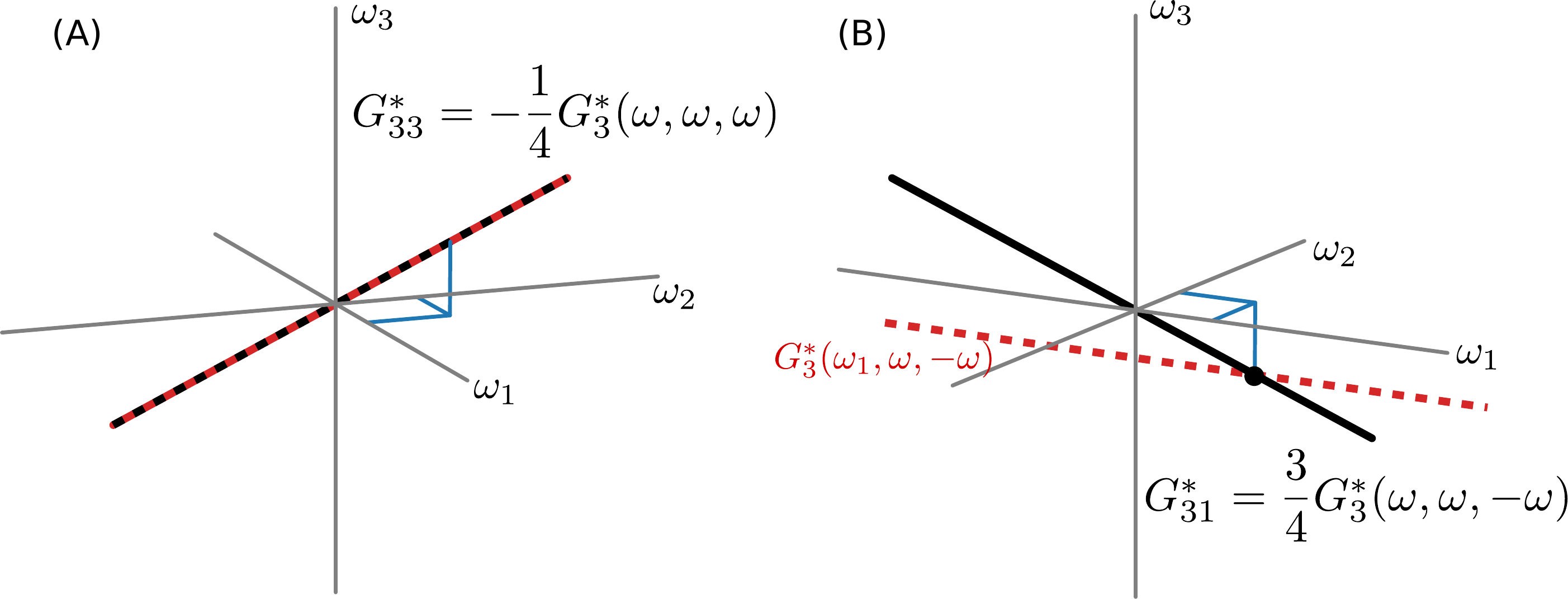}
    \caption{Schematic diagram illustrating KKR MAOS within the context of MAPS: (A) for $G_{33}^{*}$, application of  KKR at a point (blue co-ordinates) involves an integral (dashed red line) that coincides with $G_{33}^{*}$ at all frequencies; (B) for $G_{31}^{*}$ the corresponding integral involves a quantity (dashed red line) that does not coincide with $G_{31}^{*}$.}
    \label{fig:maos_schematic}
\end{figure}

Interestingly, while we can write specific expressions relating the real and and imaginary parts of $G_{33}^{*}(\omega)$ (equation \ref{eqn:kk_maos}), MAOS KKR relating the real and imaginary parts of $G_{31}^{*}(\omega)$ do not exist. In equation \ref{eqn:kk_nonlinear}, since $p_i \geq 0$, the integrand on the RHS cannot be expressed as $\eTst(u, u, -u)$, which is equal to $\eta_{31}^{*}(u)$. However, by using $p_1 = 1, p_2 = 0, p_3 = 0$ and $\omega_{1}=0, \omega_{2}=\omega, \omega_{3}=-\omega$, we obtain $\eTst(u, \omega, -\omega)$ with two fixed input frequencies in the integrand, leading to:
\begin{align}
\eta_{31}^{*}(\omega) & = \eta_{3}^{*}(\omega ,\omega ,-\omega ) = \dfrac{i}{\pi} \int\limits_{-\infty }^{\infty} \dfrac{\eta_{3}^{*}(u,\omega ,-\omega)}{u-\omega } du \notag\\
G_{31}^{*}(\omega ) & = G_{3}^{*}(\omega ,\omega ,-\omega) = \dfrac{i \omega }{\pi} \int\limits_{-\infty }^{\infty } \dfrac{G_{3}^{*}(u,\omega ,-\omega )/u}{u - \omega}du.
\label{eqn:kk_G31s}
\end{align}
The same expression can also be obtained by using $\omega_{1}=\omega$ and $\omega_{3}=-\omega$ in equation \ref{eqn:kk_maps}. This is the closest we can approach a KKR involving the MAOS modulus $G_{31}^{*}(\omega)$. This situation arises because MAOS moduli are a projection or subspace of the MAPS modulus $G_{3}^{*}(\omega_1, \omega_2, \omega_3)$. As illustrated in figure \ref{fig:maos_schematic}, the MAOS moduli can be visualized as two particular diagonal vectors (marked by black lines) in the three-dimensional domain of $G_{3}^{*}$. For $G_{33}^{*}$, the integrand of the corresponding KKR relation (equation \ref{eqn:kk_maos}) lives in the same subspace shown by the dashed red line in figure \ref{fig:maos_schematic}A. Unfortunately, the integrand of equation \ref{eqn:kk_G31s} shown by the dashed red line in figure \ref{fig:maos_schematic}B lives in a different subspace and does not yield a KKR.

\begin{table}
\begin{center}
\begin{tabular}{ccl}
\hline
\multicolumn{3}{c}{\textbf{general nonlinear KKR}}\\
\multicolumn{3}{c}{$\eta^{*}_{n}(\omega_1 + p_1 \omega, \cdots, \omega_n + p_n \omega) = \dfrac{i}{\pi}  \int\limits_{-\infty}^{\infty} \dfrac{\eta_{n}^{*}(\omega_1 + p_1 u, \cdots, \omega_n + p_n u)}{u - \omega} du, \quad p_i \geq 0$}\\[2ex]
\hline \\[-3ex]
\multirow{2}{*}{general} & $\eta_{n}^{*}$ & $\eta^{*}_{n}(\omega_1, \cdots , \omega_j, \cdots, \omega_n) = \dfrac{i}{\pi}  \int\limits_{-\infty}^{\infty} \dfrac{\eta^{*}_{n}(\omega_1, \cdots , \omega_j^{\prime}, \cdots, \omega_n)}{\omega_j^{\prime} - \omega_j} d \omega_j^{\prime}$\\ 
& $G_{n}^{*}$ & $\dfrac{G^{*}_{n}(\omega_1, \cdots , \omega_j, \cdots, \omega_n)}{\omega_j} = \dfrac{i}{\pi}  \int\limits_{-\infty}^{\infty} \dfrac{G^{*}_{n}(\omega_1, \cdots , \omega_j^{\prime}, \cdots, \omega_n)/\omega_j^{\prime}}{\omega_j^{\prime} - \omega_j} d \omega_j^{\prime}$\\  [2ex]
\hline \\[-3ex]
\multirow{2}{*}{MAPS} & $\eta_{3}^{*}$ &  $\eTst(\omega_1, \omega_2, \omega_3) = \dfrac{i}{\pi}  \int\limits_{-\infty}^{\infty} \dfrac{\eTst(\omega_1, \omega_2^{\prime}, \omega_3)}{\omega_2^{\prime} - \omega_2} d\omega_2^{\prime}$\\ 
& $G_{3}^{*}$ & $\dfrac{G^{*}_{3}(\omega_1, \omega_2, \omega_3)}{\omega_2}  = \dfrac{i}{\pi}  \int\limits_{-\infty}^{\infty} \dfrac{G^{*}_{3}(\omega_1, \omega_2^{\prime}, \omega_3)/\omega_j^{\prime}}{\omega_2^{\prime} - \omega_2} d \omega_2^{\prime}$\\  [2ex]
\hline \\[-3ex]
\multirow{2}{*}{MAOS} & $\eta_{33}^{*}$ & $\eta_{33}^{*}(\omega) = \dfrac{i}{\pi}  \int\limits_{-\infty}^{\infty} \dfrac{\eta_{33}^{*}(u)}{u - \omega} du$\\ 
& $G_{33}^{*}$ & $G_{33}^{*}(\omega) = \dfrac{i}{\pi} \omega^3 \int\limits_{-\infty}^{\infty} \dfrac{1}{u^3} \dfrac{G_{33}^{*}(u)}{u - \omega} du$ \\ [2ex] 
\hline
\end{tabular}
\end{center}
\caption{Summary of nonlinear KKR for general ($n^\text{th}$ order), MAPS and MAOS (third order) complex moduli. These equations can be derived from the general nonlinear KKR listed at the top of the table, which is derived in section \ref{sec:generalderivation}. \label{tab:nonlinear}}
\end{table}

\section{Validation of KKR for MAPS and MAOS}
\label{sec:applications}

The various KKR expressions developed hitherto are tabulated in Table \ref{tab:nonlinear} for convenience. In this section, we verify the KKR expressions corresponding to the MAPS (equation \ref{eqn:kk_maps}) and MAOS moduli (equations \ref{eqn:kk_maos} and \ref{eqn:kk_G31s}) for the single mode Giesekus model. For this model, analytical expressions for the MAPS and MAOS moduli are available in the literature \cite{Lennon2020, KateGurnon2012, Bharadwaj2015}. 

\subsection{KKR for MAPS}

The single mode Giesekus model has three parameters, the two linear parameters: modulus $G_0$, and the relaxation time $\tau_0$, and the nonlinear parameter $\alpha_G$. The zero shear viscosity is related to the linear parameters via $\eta_0 = G_0 \tau_0$. Lennon et al. \cite{Lennon2020} derived the third-order complex modulus for various constitutive equations including the single mode Giesekus model, which can be written as,
\begin{equation}
\dfrac{\eta_{3}^{*} ( \omega_{1},\omega_{2}, \omega_{3})}{\eta_{0} \tau_{0}^{2}} =\dfrac{\alpha_{G} \left( ( 3-2\alpha_{G} ) + i \sum\limits_{j}{{z_{j}}} \right)}{3\left( \prod\limits_{j}{\left( 1+i  z_{j} \right)} \right)}\dfrac{\left[ -3 - 4i  \sum\limits_{j}z_{j} + \sum\limits_{j} z_{j}^{2} + 3 \sum\limits_{j} \left(\prod\limits_{k \ne j}z_{k}\right) \right]}{\left[ \prod\limits_{j}\left( 1+i \sum\limits_{k\ne j}{{z_{k}}} \right) \right]\left( 1 + i  \sum\limits_{j} z_{j} \right)},
\label{eqn:MAPS_Gie}
\end{equation}
using the dimensionless frequency $z_i = \omega_i \tau_0$, with $i = 1, 2, 3$, for brevity. To validate the MAPS KKR (equation \ref{eqn:kk_maps}), we consider the integral, 
\begin{align}
\dfrac{1}{\eta_0 \tau_0^2} \int_{-\infty }^{\infty} \dfrac{\eta_{3}^{*}\left( \omega_{1}, u, \omega_{3} \right)}{\left( u-{\omega_{2}} \right)} & du  = \int_{-\infty }^{\infty } \dfrac{\alpha_{G}}{3(z- z_{2})} \dfrac{\left( 3-2\alpha_{G}  \right)+i\left( {z_{1}}+z+{z_{3}} \right)}{\left( 1 + i z_{1} \right)\left(1 + iz \right)\left( 1 + i z_{3} \right)} \times \notag\\
 & \dfrac{-3 - 4i(z_1 + z + z_3) + \left( z_{1}^{2}+{{z}^{2}}+z_{3}^{2} \right) + 3\left( z_{1}z + z z_{3} + z_{3} z_{1} \right) }{\left(1 + i\left( z_{1} + z \right) \right)\left(1 + i\left( z + z_{3} \right) \right)\left(1 + i\left( z_{3}+ z_{1} \right) \right)\left(1 + i\left( z_{1}+z+ z_{3} \right) \right)} dz,
\label{eqn:MAPS_Gie_KK}
\end{align}
where $z = \tau_0 u$. This integral over $z$ can be evaluated analytically. As expected from the MAPS KKR equation \ref{eqn:kk_maps}, it leads to original expression for $\eta_{3}^{*}(\omega_{1}, \omega_{2}, \omega_{3})$ given by equation \ref{eqn:MAPS_Gie} with the appropriate prefactors. It is perhaps useful to point out that in trying to verify the MAPS KKR, a small typo was discovered in equation 61 and D8 of ref. \cite{Lennon2020} (the authors reported missing a factor of $i$ in the numerator of the first term on the RHS, which is fixed in equation \ref{eqn:MAPS_Gie}). Note that verification of the MAPS KKR automatically validates equation \ref{eqn:kk_G31s} for $\eta_{31}^{*}$, which, as alluded to before, is strictly not a KKR as it does not relate the real and imaginary parts of the same property through an integral transform.

It is known that for inelastic constitutive equations such as generalized Newtonian fluids, the linear elastic moduli at all frequencies is ${G}'=0$. Consequently, generalized Newtonian fluids do not obey the Fourier transform relation between $\Gpp$ and $G(t)$  given by equation \ref{eqn:ft}. They violate the principle of causality, and even linear KKR given by equation \ref{eqn:kk_saos_complex} do not apply. For a generalized Newtonian fluid $\eta_{3}^{*}({\omega_{1}},{\omega_{2}},{\omega_{3}})= \eta^{\prime}(0)$ = constant \cite{Lennon2020}. For this case, we get $\int_{-\infty }^{\infty } \eta_{3}^{*}\left( {\omega_{1}},u,{\omega_{3}} \right)/\left( u-{\omega_{2}} \right) du = 0$, and hence MAPS KKR does not hold either. This result is  expected because a generalized Newtonian fluid is an idealization; no real fluid demonstrates $\Gp = 0$ at all frequencies.

\subsection{KKR for MAOS: Finite Frequency Window}

Unlike $\eta_{31}^{*}$, the real and imaginary parts of $\eta_{33}^{*}$ are related through a KKR (equation \ref{eqn:kkr_eta33}). As with $\eta_{31}^{*}$, the validity of the corresponding specialized KKR is automatically implied by the validity of the MAPS KKR. MAOS KKR (equation \ref{eqn:kk_maos}) may can also be directly verified using the relations for $G_{33}^{*}$ derived by Gurnon and Wagner \cite{KateGurnon2012}, and Bharadwaj and Ewoldt \cite{Bharadwaj2015}. We obtain the same expressions from the MAPS relation (equation \ref{eqn:MAPS_Gie}) using $\omega_{1} = \omega_{2} = \omega_{3} = \omega$,
\begin{align}
\dfrac{G_{33}^{\prime}(\omega)}{G_0} = -\dfrac{1}{4} \dfrac{G_{3}^{\prime}(\omega ,\omega ,\omega)}{G_0} & = \dfrac{\alpha_{G}  z^4  (-21 + 30z^2 + 51z^4 + 4  \alpha_{G}  (4 - 17z^2 + 3z^4))}{4(1 + z^2)^3  (1 + 4z^2)  (1 + 9z^2)} \notag \\
\dfrac{G_{33}^{\prime\prime}(z)}{G_0} = -\dfrac{1}{4} \dfrac{G_{3}^{\prime\prime}(\omega ,\omega ,\omega)}{G_0} & = \dfrac{\alpha_{G}  z^3  (-3  + 48z^2 + 33z^4 -18z^6 + \alpha_G  (2 - 48z^2 + 46z^4))}{4(1 + z^2)^3  (1 + 4z^2)  (1 + 9z^2)},
\label{eqn:maos_giesekus}
\end{align}
with $z = \omega \tau_0$. Note that there is a small typo in Bharadwaj and Ewoldt \cite{Bharadwaj2015} in the expression for $G_{33}^{\prime}$ where the term $+30z^2$ in the numerator is accidentally replaced by $-30z^2$.

Note that MAOS KKR require knowledge of $G_{33}^{*}$ over an infinite frequency window. In what follows, we explore the sensitivity of MAOS KKR when experimental data are available only over a limited frequency range, $\omega_{\min} \leq u \leq \omega_{\max}$. We denote these approximations of MAOS KKR by decorating the corresponding moduli with a tilde,
\begin{figure}
\begin{center}
\includegraphics[scale=0.6]{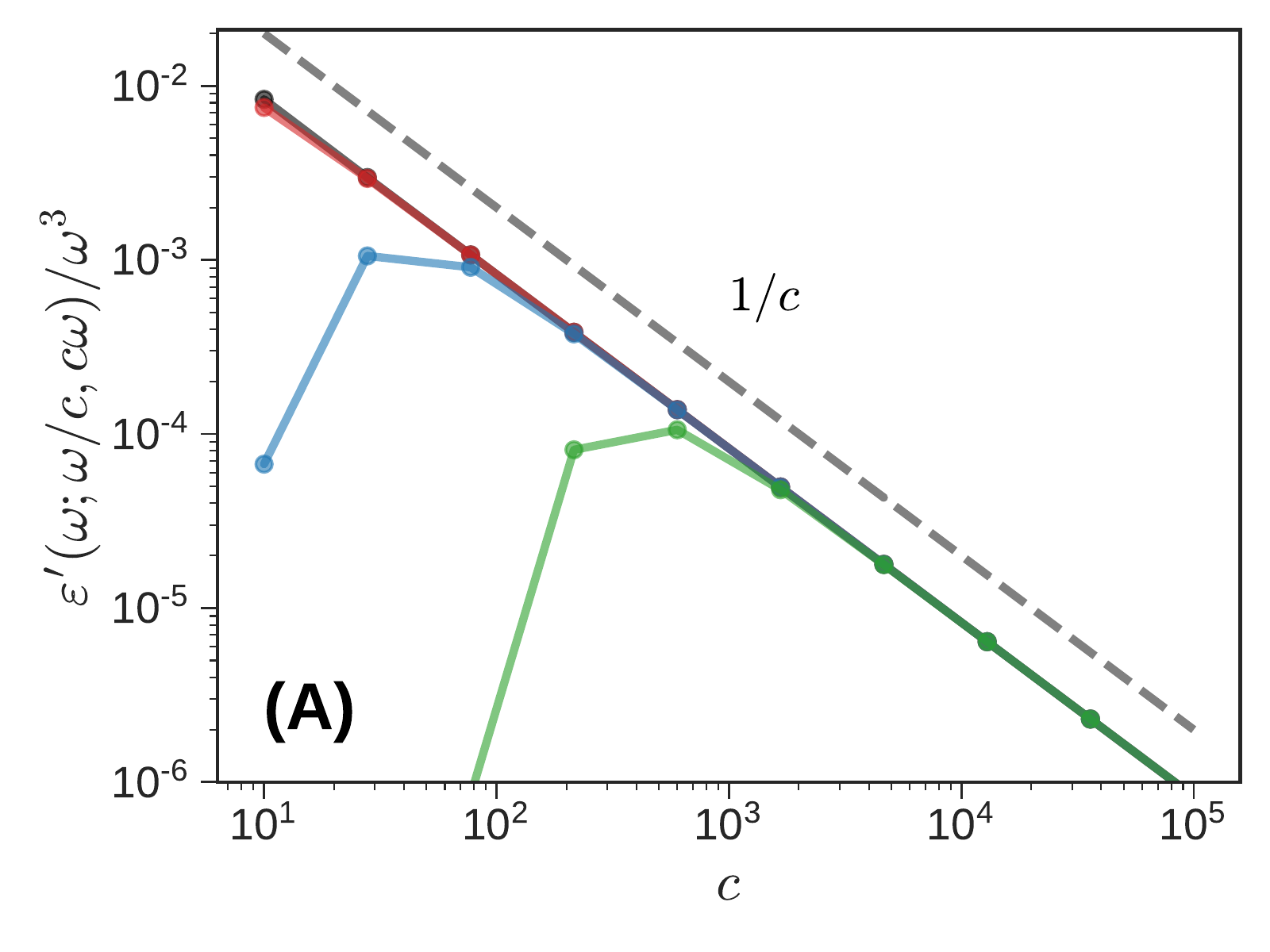}\\
\includegraphics[scale=0.6]{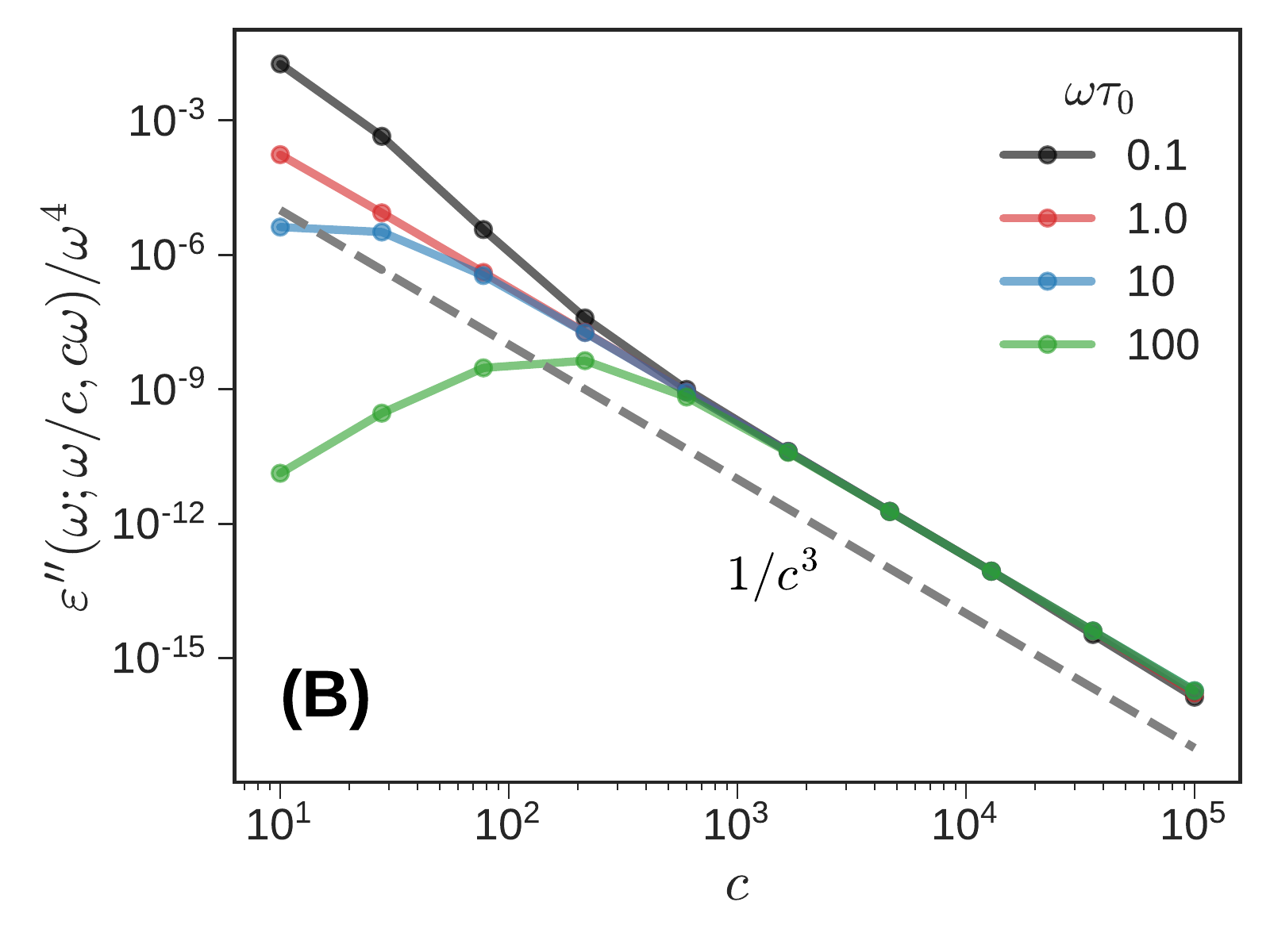}
\caption{Normalized errors for finite-frequency approximations of MAOS KKR for the single mode Giesekus model with $G_0 = \tau_0 = 1$ and $\alpha_{G}=0.2$: (A) $\varepsilon_{33}^{\prime}/\omega^3$ and (B) $\varepsilon_{33}^{\prime\prime}/\omega^3$ are plotted as a function of the parameter $c$, which controls the width of the frequency window via $\omega_{\min} = \omega/c$  and $\omega_{\max} = c\omega$. Four different values of the frequency $\omega \tau_{0}$ are selected. When the low frequency correction dominates the error, we expect $\varepsilon_{33}^{\prime} \propto\omega^3/c$ and $\varepsilon_{33}^{\prime\prime} \propto\omega^4/c^3$, which are indicated by the dashed gray lines.\label{fig:errorc}}
\end{center}
\end{figure}
\begin{align}
\tilde{G}_{33}^{\prime }(\omega ) & = -\dfrac{2\omega^{4}}{\pi} \int_{\omega_{\min}}^{\omega_{\max}} \dfrac{1}{u^{3}} \dfrac{G_{33}^{\prime\prime}(u)}{\left(u^{2}- \omega^{2} \right)}du \notag\\
\tilde{G}_{33}^{\prime \prime }(\omega) & = \dfrac{2\omega^{3}}{\pi} \int_{\omega_{\min}}^{\omega_{\max}} \dfrac{1}{{{u}^{2}}}\dfrac{G_{33}^{\prime }(u)}{\left( u^{2}- \omega^{2} \right)}du.	
\label{eqn:kk_maos_approx}
\end{align}
Fortunately, these integrals can be evaluated analytically for the Giesekus model, although the resulting expressions are somewhat elaborate. We can define the absolute error between the true and approximate moduli to quantify the sensitivity of KKR to truncation of the frequency window as,
\begin{align}
\epsilon_{33}^{\prime}(\omega; \omega_{\min}, \omega_{\max}) & = \left| G_{33}^{\prime }(\omega) - \tilde{G}_{33}^{\prime}(\omega; \omega_{\min}, \omega_{\max}) \right| \notag\\
\epsilon_{33}^{\prime\prime}(\omega; \omega_{\min}, \omega_{\max}) & = \left| G_{33}^{\prime\prime }(\omega) - \tilde{G}_{33}^{\prime\prime}(\omega; \omega_{\min}, \omega_{\max}) \right|.
\label{eqn:kk_error}
\end{align}

The error due to truncation has contributions from the left, $u \in [0, \omega_{\min}]$, and right, $u \in [\omega_{\max}, \infty]$, tails. At large frequencies $u \rightarrow \infty$, the integrands of both the MAOS KKR decay rapidly as $1/u^6$,
\begin{equation}
\lim_{u \rightarrow \infty}  \dfrac{1}{u^{3}} \dfrac{G_{33}^{\prime\prime}(u)}{\left(u^{2}- \omega^{2} \right)}
\propto \dfrac{1}{u^6}, \quad \lim_{u \rightarrow \infty} \dfrac{1}{{{u}^{2}}}\dfrac{G_{33}^{\prime }(u)}{\left( u^{2}- \omega^{2} \right)} \propto \dfrac{1}{u^6}.
\end{equation}
Thus, the typical correction due to truncation of the high frequency or right tail is relatively modest compared to the truncation of the left tail, which we consider next. At low frequencies,
\begin{equation}
\lim_{u \rightarrow 0}  \dfrac{1}{u^{3}} \dfrac{G_{33}^{\prime\prime}(u)}{\left(u^{2}- \omega^{2} \right)}
\propto \dfrac{u^{0}}{\omega^2}, \quad \lim_{u \rightarrow 0} \dfrac{1}{{{u}^{2}}}\dfrac{G_{33}^{\prime }(u)}{\left( u^{2}- \omega^{2} \right)} \propto \dfrac{u^2}{\omega^2}.
\end{equation}
This left tail contribution to the error in equation \ref{eqn:kk_error} can be approximated as,
\begin{align}
\epsilon_{33}^{\prime}(\omega; \omega_{\min}, \infty) & = -\dfrac{2\omega^{4}}{\pi} \int_{0}^{\omega_{\min}} \dfrac{1}{u^{3}} \dfrac{G_{33}^{\prime\prime}(u)}{\left(u^{2}- \omega^{2} \right)} du \approx -\dfrac{2\omega^{4}}{\pi} \int_{0}^{\omega_{\min}} \dfrac{u^{0}}{\omega^2} du \propto \omega^2 \omega_{\min} \notag \\
\epsilon_{33}^{\prime\prime}(\omega; \omega_{\min}, \infty) & = \dfrac{2\omega^{3}}{\pi} \int_{0}^{\omega_{\min}} \dfrac{1}{u^{2}} \dfrac{G_{33}^{\prime}(u)}{\left(u^{2}- \omega^{2} \right)} du \approx \dfrac{2\omega^{3}}{\pi} \int_{0}^{\omega_{\min}} \dfrac{u^{2}}{\omega^2} du \propto \omega \omega_{\min}^3. 
\label{eqn:errorAnal}
\end{align}

Figure \ref{fig:errorc} depicts the total truncation error (equation \ref{eqn:kk_error}) at four different frequencies $\omega \tau_0 =  $ 0.1, 1, 10, and 100 as a function of the width of the frequency window, which is controlled by the parameter $c$. We set $\omega_{\min} = \omega/c$ and $\omega_{\max} = c\omega$ for each choice of $\omega$ and $c$. As $c \rightarrow \infty$, the frequency window becomes infinite, and the errors $\varepsilon_{33}^{\prime}$ and $\varepsilon_{33}^{\prime\prime}$ go to zero. Error analysis (equation \ref{eqn:errorAnal}), which assumes that the left tail is primarily responsible suggests, $\varepsilon_{33}^{\prime} \propto \omega^2 (\omega/c) = \omega^3/c$. This is clearly evident in figure \ref{fig:errorc}A at sufficiently large $c$, where the $1/c$ dependence is shown by the dashed line, and the error is normalized by $\omega^3$. Similar analysis suggests that $\varepsilon_{33}^{\prime\prime} \propto \omega (\omega/c)^3 = \omega^4/c^3$, which is also evident at large $c$ in figure \ref{fig:errorc}B.

At smaller values of $c$ deviations from the asymptotic trendlines are visible. These deviations surface when the finite frequency window $[\omega/c, c\omega]$ does not include a sufficient information around the characteristic relaxation time of the Giesekus model, $\tau_0$. That is, it is important for the finite frequency window to include sufficient information around the corresponding characteristic frequency, i.e. $\omega_{\min} \ll 2\pi/\tau_0 \ll \omega_{\max}$. The dataseries corresponding to for $\omega \tau_0 = 1$ always includes this region for the range of $c$ explored in \ref{fig:errorc}. Thus, it tracks the asymptotic trendline more faithfully than other frequencies.

\begin{figure}
\begin{center}
\includegraphics[scale=0.6]{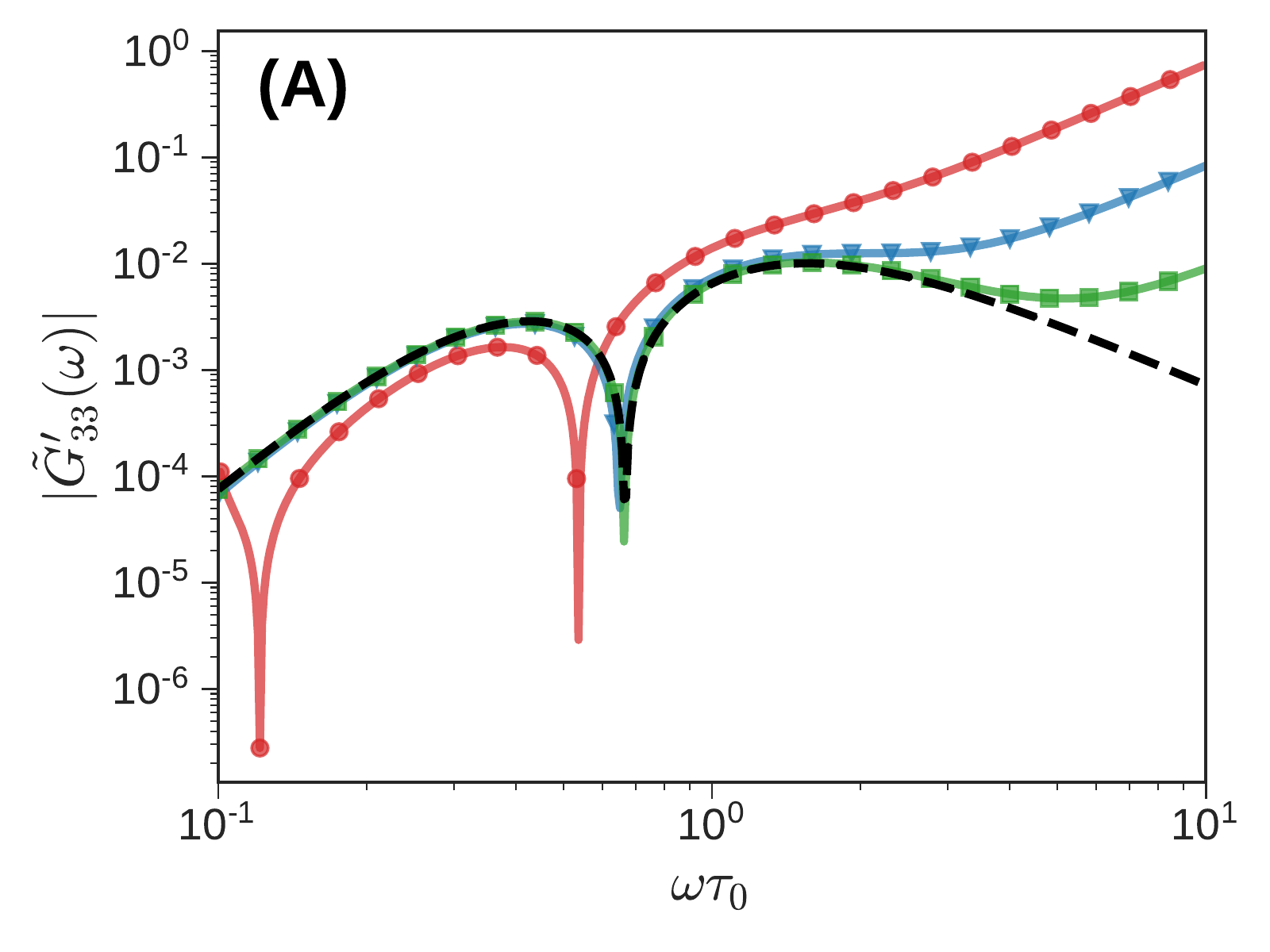}\\
\includegraphics[scale=0.6]{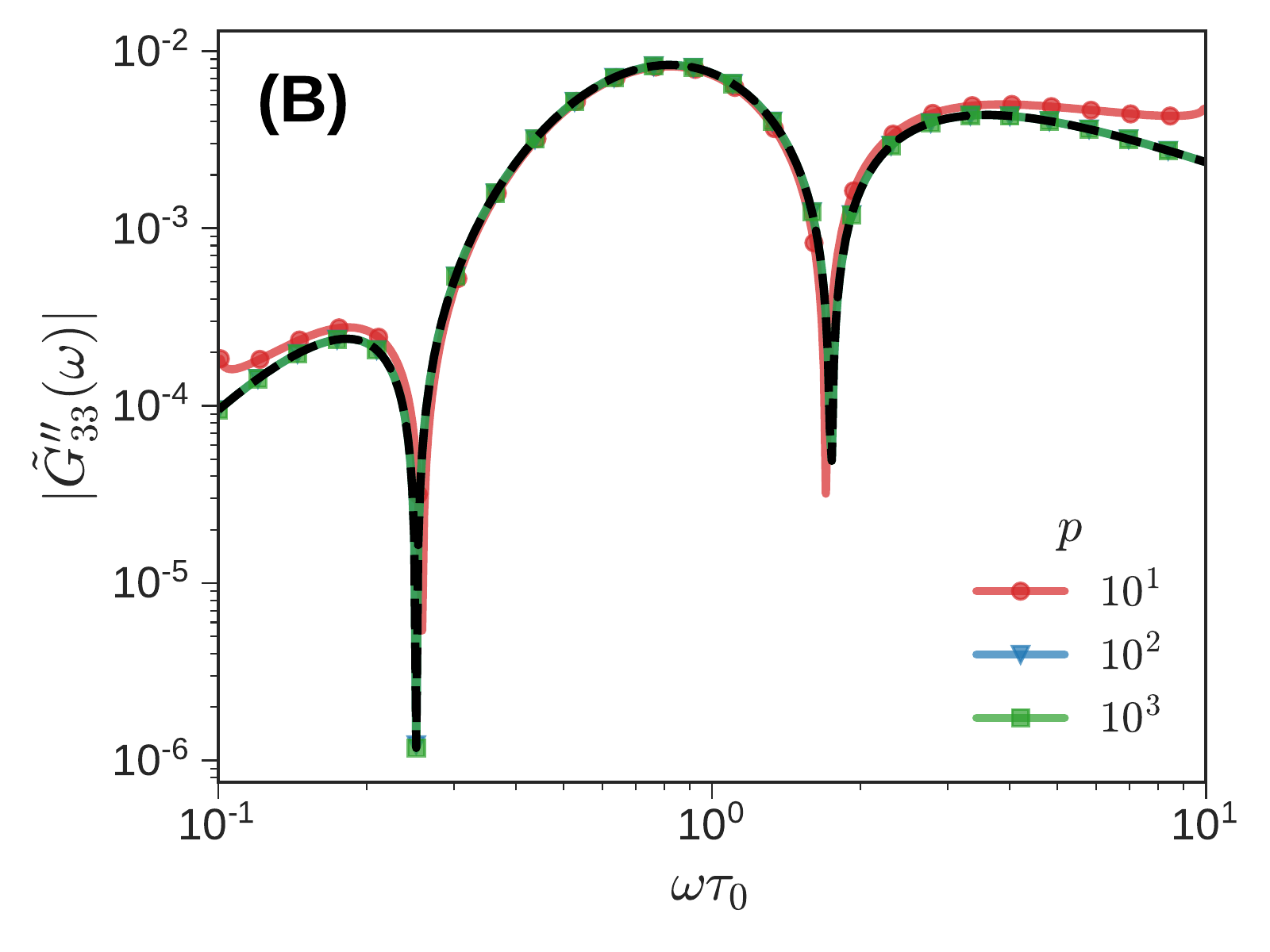}
\caption{Finite-frequency window approximations of the third harmonic MAOS moduli $\tilde{G}_{33}^{\prime }(\omega)$ and $\tilde{G}_{33}^{\prime \prime}(\omega)$ for the single mode Giesekus model with $G_0 = \tau_0 = 1$ and $\alpha_{G}=0.2$. Three different widths of the frequency window where $\omega_{\min} \tau_0 = 1/p$  and $\omega_{\max} \tau_0  = p$ are shown. The dashed black lines are the true MAOS moduli. \label{fig:errorp}}
\end{center}
\end{figure}

Two other practical observations can be made from the figure: (i) at a given $\omega$ and $c$, the magnitude of the error $\varepsilon_{33}^{\prime\prime}$ is smaller, often much smaller, than the corresponding error $\varepsilon_{33}^{\prime}$, and (ii) for a fixed but sufficiently large frequency window, both errors increase rapidly with frequency $\omega$. This suggests that limiting the frequency window adversely affects high frequency predictions of $\tilde{G}_{33}^{\prime}(\omega)$ and $\tilde{G}_{33}^{\prime \prime}(\omega)$.

In practice, standard rheometers have a fixed frequency range, typically between $10^{-3}$ to $10^3$ rad/s. Hence, we can explore the sensitivity of $\tilde{G}_{33}^{\prime}(\omega)$ and $\tilde{G}_{33}^{\prime \prime}(\omega)$, when the experimental data are available from a fixed frequency window $\omega_{\min} = 1/p$ and $\omega_{\max} = p$. Figure \ref{fig:errorp} shows $\tilde{G}_{33}^{\prime}(\omega)$  for $\omega \tau_0 \in [0.1, 10]$, at three different values of $p$ = 10, 100, and 1000.

As expected, the agreement between the approximate and true moduli improves as $p$ increases. Note that when $p = 10$, the prediction, particularly for $\tilde{G}_{33}^{\prime}(\omega)$, is poor. The corresponding prediction for $\tilde{G}_{33}^{\prime\prime}(\omega)$ is not quite as bad. This seems to be a general trend: predictions of 
$G_{33}^{\prime\prime}(\omega)$ using MAOS KKR on limited frequency data for $G_{33}^{\prime}(\omega)$ are far more reliable than vice versa. This was foreshadowed by figure \ref{fig:errorc}B, where the error $\varepsilon^{\prime\prime} < \varepsilon^{\prime}$ at the same value of $\omega$ and $c$. 

However, even for $G_{33}^{\prime\prime}(\omega)$ we need data approximately one order of magnitude larger ($p =100$ corresponds to $\omega_{\min} = 10^{-2}$ and $\omega_{\max} = 10^{2}$) than the range of reliable prediction ($\omega \tau_0 \in [0.1, 10]$) shown in figure \ref{fig:errorp}B. Increasing $p$ by another order of magnitude to 1000, results only in minor improvements in prediction of $G_{33}^{\prime\prime}(\omega)$. Unlike $G_{33}^{\prime\prime}(\omega)$, the predictions of $\tilde{G}_{33}^{\prime}(\omega)$ at high frequency, even at $p = 1000$ are quite poor.  This is anticipated by equation \ref{eqn:errorAnal}, which suggested that the error for the storage modulus is proportional to $\omega^2/p$, unlike the loss modulus which is proportional to $\omega/p^3$ and much better behaved.

\section{Summary}

Linear KKR are integral transforms that relate the real and imaginary parts of the complex modulus $G^{*}$ (or complex viscosity $\eta^{*}$). These relations are a mathematical reflection of the principle of causality which constrains the linear relaxation modulus ($G(t < 0) = 0$). We started with a multiple integral expansion that generalizes the Boltzmann superposition principle to nonlinear rheology. Nonlinear KKR, similar to their linear counterparts, arises from the principle of causality which also constrains $n^\text{th}$ order relaxation modulus $G_n(t_1, \cdots, t_n)$.

We derived a general form of nonlinear KKR (equation \ref{eqn:kk_nonlinear}) following the approach of Hutchings et al. \cite{Hutchings1992}. We specialized this general KKR to MAPS rheology which relates the real and imaginary parts of the third-order complex modulus $G_3^{*}(\omega_1, \omega_2, \omega_3)$ or complex viscosity $\eta_3^{*}(\omega_1, \omega_2, \omega_3)$ (equation \ref{eqn:kk_maps}). Recall that knowledge of $G_3^{*}(\omega_1, \omega_2, \omega_3)$ allows us to predict the asymptotically nonlinear material response to any arbitrary medium amplitude deformation history. MAOS rheology can then be considered as a popular special case of MAPS rheology that is characterized by two moduli $G_{31}^{*} = (3/4) G_3^{*}(\omega, -\omega, \omega)$ and $G_{33}^{*} = (-1/4) G_3^{*}(\omega, \omega, \omega)$. While a MAOS KKR relating the real and imaginary parts of $G_{33}^{*}$ can be written, no such expression relating the the real and imaginary parts of $G_{31}^{*}$ exists.

We verified the MAPS KKR relations on the single mode Giesekus model for which the third-order complex modulus $G_3^{*}(\omega_1, \omega_2, \omega_3)$ is analytically known. With practical applications in mind, we  investigated the sensitivity of the MAOS KKR when the domain of integration is truncated and data is limited to a finite frequency window. We found that: (i) the truncation error is typically dominated by the low-frequency or left tail, (ii) inferring $G_{33}^{\prime\prime}$ from $G_{33}^{\prime}$ is more reliable than vice versa, (iii) making predictions over a particular frequency range requires approximately an extra decade of data beyond the frequency range of prediction, and (iv) predictions of $G_{33}^{\prime}$ at large frequencies are poor, even when two decades of data beyond the prediction range are available.

\section*{Acknowledgments}

This work is based in part upon work supported by the National Science Foundation under grant no. NSF DMR-1727870 (SS). YMJ acknowledges financial support from Science and Engineering Research Board (SERB), Department of Science and Technology, Government of India. The authors thank Kyle R. Lennon and Gareth M. McKinley for help with equation \ref{eqn:MAPS_Gie}, and Ms. Shweta Sharma for assistance with Mathematica.

\section*{Data Availability Statement}
Data that support the findings of this study are available from the corresponding author upon reasonable request.

\bibliography{extracted1}

\end{document}